\documentclass{ifacconf}

\usepackage{graphicx}      % include this line if your document contains figures
\usepackage{natbib}        % required for bibliography
\usepackage{IRM_nothm}
\begin{document}
\begin{frontmatter}

\title{On Existence of Separable Contraction Metrics 
for Monotone Nonlinear Systems\thanksref{footnoteinfo}} 

\thanks[footnoteinfo]{This work was supported by the Australian Research Council.}

\author[First]{Ian R. Manchester} 
\author[Second]{Jean-Jacques E. Slotine}

\address[First]{Australian Centre for Field Robotics, School of Aerospace, Mechanical and Mechatronic Engineering, University of Sydney (e-mail:ian.manchester@sydney.edu.au).}
\address[Second]{Nonlinear Systems Laboratory, Massachusetts Institute of Technology (e-mail:jjs@mit.edu)}

\begin{abstract}                % Abstract of not more than 250 words.
Finding separable certificates of stability is important for tractability of analysis methods for large-scale networked systems. In this paper we consider the question of when a nonlinear system which is contracting, i.e. all solutions are exponentially stable, can have that property verified by a separable metric. Making use of recent results in the theory of positive linear systems and separable Lyapunov functions, we prove several new results showing when this is possible, and discuss the application of to nonlinear distributed control design via convex optimization.
\end{abstract}


\end{frontmatter}
%===============================================================================

\section{Introduction}

Many emerging applications of system analysis and control involve
large networks of interconnected nonlinear dynamic systems: smart-grid
power systems, traffic management with autonomous vehicles, internet
congestion control, and analysis of biological signalling networks, to
name but a few. In order to make analysis methods scalable to large
networks and robust to node dropouts or additions, it is crucial to be
able to understand the overall network behaviour by way of conditions
on just the local node dynamics and their interactions with immediate
neighbours.

The traditional method for stability analysis makes use of a Lyapunov
function: a positive-definite function of the system's state that
decreases under flows of the system. Finding Lyapunov functions for
arbitrary large-scale systems is generally intractable, but can be greatly
simplified if one restricts the search to {\em separable} Lyapunov
functions, in particular sum-separable, i.e. $V(x) = \sum_i
V_i(x_i)$, or max-separable, i.e. $V(x) = \max_i V_i(x_i)$, where
$x_i$ denotes the state of the $i^{th}$ subsystem, and $x$ denotes the
concatenation of all subsystem states \citep{dirr2015separable}.

There has been substantial work recently on establishing when such separable Lyapunov functions should exist. In the case of linear positive systems, i.e. linear systems for which the non-negative orthant is flow-invariant, the Perron-Frobenius theory gives separable Lyapunov functions \citep{berman_nonnegative_1994}. More recently, several researchers have taken advantage of these properties to dramatically simplify problems of decentralized control design and system identification \citep{tanaka_bounded_2011}, \citep{tanaka2013symmetric}, \citep{colombino2015robust}, \citep{rantzer_scalable_2015}, \citep{umenberger_scalable_2016}. In a recent paper, fundamental results about separable Lyapunov functions for positive systems have been extended to linear time-varying systems \citep{khong_diagonal_2016}.

For nonlinear systems . Recent work has focused on establishing conditions on existence of separable Lyapunov functions, especially in connection with input-to-state stability properties \cite{dashkovskiy_small_2010}, \cite{ito_capability_2012} and monotone systems \cite{dirr2015separable}, the natural nonlinear generalization of a linear positive system \cite{smith_monotone_1995}

Contraction analysis generalizes techniques from linear systems to nonlinear systems. Roughly speaking, a system is contracting if the linearization along {\em every} solution is exponentially stable \cite{Lohmiller98}. This property is verified by the existence of a {\em metric}, which can be taken to be  of Riemannian form though others are possible. This idea can be traced back to \cite{lewis1949metric} and has been explored in more detail recently by \cite{forni2014differential} and extended to analysis of limit cycles by \cite{manchester2013transverse}. Advantages of contraction analysis include the fact that questions of stability are decoupled from knowledge of a particular solution, unlike Lyapunov theory, and because the object of study is a family of linear time-varying systems, many familiar results can be extended to nonlinear systems. 

A natural question to ask is when these contraction conditions can be verified in a scalable way. It was noted in \cite{coogan_separability_2016} that some commonly used metrics based on $l^1$ and $l^\infty$ norm are naturally separable, a property which was used implicitly in several papers on networked system analysis and control, e.g. \citep{russo2011graphical, como2015throughput}. However, this does not answer the question when such a separable metric exists for systems which are known to be contracting with respect to a (not necessarily separable) metric, as in the case of linear positive systems.

 %Basic results on contraction Can be traced back to the numerical analysis literature~\cite{lewis1949metric,Hartmann,Demidovich}.
Recently, the technique of control contraction metrics has been introduced, which extends contraction analysis to constructive control design \citep{manchester_control_2015}. In fact, synthesis conditions can be transformed to a convex problem: a set of pointwise linear matrix inequalities, which can be verified using sum-of-squares programming (see e.g. \cite{aylward2008stability}). Despite being convex, these conditions do not generally scale to very large systems.

It was recently shown by \cite{shiromoto_decentralized_2016} that if one restricts the search to sum-separable control contraction metrics, then the problem of distributed control synthesis can be made convex. This can be considered an extension of the results of \cite{tanaka_bounded_2011} to a class of nonlinear systems.

Another motivating application is nonlinear  system identification with guaranteed model stability, building on the results of \cite{tobenkin2010convex}. Recent results allow scalable computation for linear systems \citep{umenberger_scalable_2016}. With separable contraction metrics, these could be extended to identification of large-scale nonlinear systems. 

\section{Notation and Preliminaries}

For symmetric matrices $A,B$ the notation $A\ge B$ ($A>B$) means that $A-B$ is positive semidefinite (positive definite), whereas for vectors $x,y \in \RR^n$, the notation $x\ge y$ denotes element-wise inequality.
The non-negative reals are denoted $\RR^+:=[0,\infty)$, and the natural numbers from 1 to $n$ are denoted $\N_{1,n}$. A smooth matrix function $M(x,t)$ is called {\em uniformly bounded} if there exists $\alpha_2\ge\alpha_1>0$ such that $\alpha_1 I\le M(x,t) \le \alpha_2 I$ for all $x,t$. Given a vector field $v:\R^n\times\R^+\to \R^n$ defined for $x\in\R^n,t\in\R^+$, we use the following notation for directional derivative of a matrix function: $\partial_v M := \sum_j\pder[M]{x_j}v_{j}$. 

In this paper we consider time-varying nonlinear systems:
\begin{equation}\label{eq:sys}
\dot x = f(x,t)
\end{equation}
where $x(t)\in\RR^n$ is the state vector at time $t\in\RR^+$, and $f$ is a smooth function of $x$ and at least piecewise-continuous in $t$, though these can be relaxed somewhat. Note that this system representation can include systems with external control or disturbance inputs $\dot x(t) = f(x(t),u(t),w(t))$, where for our purposes we absorb these into the time-variation in \eqref{eq:sys}.

A dynamical system is {\em monotone} if for any pair of solutions $x^a$ and $x^b$,  $x^a(0)\le
x^b(0)$ implies $x^a(t)\le x^b(t)$ for all $t \ge 0$, where $\le$ denotes
component-wise inequality. This property can be generalized to
partial-orderings based on arbitrary cones, however the positive
orthant is the natural cone for the purposes of studying
separability. A differential characterization of monotonicity is that
the off-diagonal elements of $\pder[f]{x}$ are non-negative -- this is
implied by the {\em Kamke-M\"uller conditions}
\citep{smith_monotone_1995}.

Internally positive linear systems form an important subset of monotone systems. A continuous-time linear system 
\[
\dot x = A(t)x
\]
is positive (and hence monotone) if $A_{ij}(t)\ge 0$ for all $i\ne j$ and for all $t$. In the case of time-invariant systems, the following result is well-known (see e.g. \cite{berman_nonnegative_1994}, \cite{rantzer_scalable_2015}).
\begin{thm}
	If $A$ is positive and Hurwitz, then there exists $p_i>0, q_i>0$, and $d_i>0$, $i = 1, 2, ..., n$  such that the following functions
	\begin{align}
	V_p(x) &= \sum_i p_i |x_i|,\\
	V_q(x) &= \max_i \{q_i |x_i|\},\\
	V_d(x) &= \sum_i d_i|x_i|^2,
	\end{align}
	are Lyapunov functions for the system $\dot x = Ax$. Moreover, one can take $d_i = p_iq_i$.
\end{thm}
Note that $V_d(x)$ is a quadratic Lyapunov function $V_d(x)=x'Dx$ for which $D$ is diagonal and $D_{ii}=d_i$, and $V_p$ is linear on the non-negative orthant.

A recent paper partially extends these results to linear time-varying (LTV) systems:
\begin{thm}[\cite{khong_diagonal_2016}]\label{thm:khong}
A linear positive system $\dot x = A(t)x$, with $A(t)$ piecewise continuous and uniformly bounded for all $t\in\RR^+$, is exponentially  stable if and only if there exists a Lyapunov function $V(x,t) = x'P(t)x$ with $P$ diagonal such that $\eta|x|^2\le V(x,t)\le \rho |x|^2$ and $\dot V(x,t) \le - \nu |x|^2$ for all $t\ge 0$ and some $\eta, \rho, \nu>0$.
\end{thm}

We utilize the following standard results of Riemannian geometry, see, e.g., \cite{docarmo1992riemannian} for details. A Riemannian metric is a smoothly-varying inner product $\ip{\cdot}{\cdot}_x$ on the tangent space of a state manifold $\X$; this defines local notions of length, angle, and orthogonality. In this paper $\X=\R^n$ and the tangent space can also be identified with $\R^n$. We allow metrics to be smoothly time-varying, and use the following notation: $\ip{\delta_1}{\delta_2}_{x,t}=\delta_1'M(x,t)\delta_2$ and $\|\delta\|_{x,t} = \sqrt{\ip{\delta}{\delta}_{x,t}}$. We call a metric {\em uniformly bounded} if $\exists \alpha_2\ge \alpha_2>0$ such that $\alpha_1 I\le M(x,t) \le \alpha_2 I$ for all $x,t$. For a smooth curve $c:[0,1]\to\R^n$ we use the notation $c_s(s) :=\pder[c(s)]{s}$, and define the Riemannian length and energy functionals as 
\[
L(c,t):=\int_0^1\|c_s\|_{c,t}ds, \ \ E(c,t):=\int_0^1\|c_s\|^2_{c,t}ds,\] 
respectively, with integration interpreted as the summation of integrals for each smooth piece. 
Let $\Gamma$ be the set of piecewise-smooth curves $[0,1]\to\R^n$, and for a pair of points $x,y\in\R^n$, let $\Gamma(x,y)$ be the subset of $\Gamma$ connecting $x$ to $y$, i.e. curves $c\in\Gamma(x,y)$ if $c\in\Gamma$, $c(0)=x$ and $c(1)=y$. A smooth curve $c(s)$ is {\em regular} if $\pder[c]{s}\ne 0$ for all $s\in[0,1]$. The Riemannian distance $d(x,y,t):=\inf_{c\in\Gamma(x,y)}L(c,t)$, and we define $E(x,y,t):=d(x,y,t)^2$. Under the conditions of the Hopf-Rinow theorem a smooth, regular minimum-length curve (a geodesic) $\gamma$ exists connecting every such pair, and the energy and length satisfy the following inequalities: $E(x,y,t)=E(\gamma,t)=L(\gamma,t)^2\le L(c,t)^2 \le E(c,t)$ where $c$ is any curve joining $x$ and $y$. For time-varying paths $c(t,s)$, we also write $c(t):=c(t,\cdot):[0,1]\to\R^n$.

A nonlinear system \eqref{eq:sys} is called contracting if {\em all} solutions are exponentially stable.  A central result of \cite{Lohmiller98} is that if there exists a uniformly bounded metric $M(x,t)$ such that
\[
\dot M + \pder[f]{x}'M+M\pder[f]{x}\le -2\lambda M,
\]
where $\dot M = \pder[M]{t}+\partial_f M$, 
then the system is contracting with rate $\lambda$. This inequality states that $\dot V \le -2\lambda V$ where $V(x,\delta,t) = \delta'M(x,t)\delta$. From this, it is straightforward to establish that the Riemannian distance (and energy) between any pair of points decreases exponentially, and thus can serve as  incremental Lyapunov functions. %We call a system {\em strictly contracting with rate $\lambda$} if $\ddt \|\delta_x\|_{x,t}< -\lambda \|\delta_x\|_{x,t}$ for $\delta_x\ne 0$. Since $M(x,t)>0$, a system which is contracting with rate $\lambda>0$ is strictly contracting with any rate less than $\lambda$, so the ``strictness'' could be thought of as an attribute of the {\em rate} more than the {\em system}.

Non-Riemannian contraction metrics, e.g. based on $l^1$ and $l^\infty$ norms, can also be used \citep{Lohmiller98} and offer simplified tests of contraction for certain applications, see, e.g. \cite{russo2011graphical}.

\section{Problem Statement}

Let the state vector $x$ of \eqref{eq:sys} be partitioned into $N$ node states $x = [x_1', x_2', ..., x_N']$ where $x_i\in \RR^{n_i}, i=1, ..., N$ and $n=\sum_{i = 1, ..., N} n_i$. The objective is to determine when contraction can be established using a set of local node metrics. In general, local nodes may have vector states although our main result will relate the the situation where $n_i=1$ for all $i$, i.e. the state of each node is a scalar.

Following the terminology from Lyapunov functions \citep{dirr2015separable}, a contraction metric $V(x,\delta)$ is called {\em sum-separable} if it can be decomposed as
\[
V(x,\delta) = \sum_{i=1}^N \delta_i' M_i(x_i)\delta_i
\]
and {\em max-separable} if 
\[
V(x,\delta) = \max_{i\in \N_{1,N}} \{\delta_i' M_i(x_i)\delta_i\}
\]
We will focus on the above (sum-type) notion of separability, but will also briefly discuss max-separable contraction metrics.

The following problem is, to the authors' knowledge, open:
\begin{prob}
	Suppose a dynamical system of the form \eqref{eq:sys} is both contracting and monotone. Characterize the additional conditions (if any) that imply the existence of a separable contraction metric.
\end{prob}

For the case of linear systems, it is known that no further conditions are required beyond contraction (exponential stability) and monotonicity (positivity). In this paper, we do not solve this problem, but rather present some partial results towards this goal.

\section{Motivation: distributed control design}

As well as producing scalable analysis conditions, a major motivation for separable contraction metrics is that they convexify the problem of distributed control design for nonlinear systems \citep{shiromoto_decentralized_2016}. 
Here we briefly recap these results and provide new results.

Consider a control system of the form
\begin{equation}\label{eq:control_sys}
\dot x = f(x)+B(x)u	
\end{equation}

 In \cite{manchester_control_2015} it was shown that if a metric exists such that the following implication is true:
\begin{equation}\label{eq:ccm}
\delta'MB=0 \Longrightarrow \delta'(\dot M + A'M + MA+2\lambda M)< 0
\end{equation}
then all solutions of \eqref{eq:control_sys} are exponentially stabilizable by state feedback. Furthermore, the Riemannian energy to a target solution is a control Lyapunov function which verifies this. By the formula for first variation,  rate of change of Riemannian energy is
\begin{align}
\frac{1}{2}\frac{d}{dt} E =& - \langle \gamma_s(t,1), f(x,t)+B(x,t)u\rangle_{x,t}
-\langle \gamma_s(t,1),\rangle_{x,t}\notag\\
& +\langle \gamma_s(t,0), \dot x^\star \rangle_{x^\star,t}+\frac{1}{2}\pder[E]{t}\label{eqn:energy_decrease}\end{align}

To be precise, \eqref{eq:ccm} implies that the set of $u$ for which \eqref{eqn:energy_decrease} is satisfied is non-empty. Such a $u$ can be found either by a path integral construction, or by solving e.g. a linear or quadratic program with the constraint \eqref{eqn:energy_decrease}.

As was demonstrated by \cite{shiromoto_decentralized_2016}, when the metric is sum-separable the computation of the controller depends only on information from  the local state information and neighbours. In this case, computation of the Riemannian energy (by way of finding a minimal geodesic between $x^\star$ and $x$) can be done a completely decentralized way, using only local node information, and a stabilizing feedback control can be constructed using only local and neighbor information.

For systems with vector nodes, i.e. $n_i>0$ for at least one $i$, computing the minimal geodesic $\gamma$ will require solving an online path optimization. Here we note in the case when nodes have scalar states this can be eliminated.

\begin{prop}
		Consider a system of the form \eqref{eq:control_sys}. Suppose a sum-separable control contraction metric exists with $n_i=1$ for all $i$. Then under the change of coordinates
	\[
z_i(x_i,t) = \int_0^{x_i}\theta_i(\xi,t)d\xi,
\]
where $\theta_i(x_i,t) = \sqrt{m_i(x_i,t)}$, 
squared Euclidean distance to any target solution $|z-z^\star|^2$ is a control Lyapunov function.
\end{prop}
\begin{pf}
Under this coordinate change, tangent vectors transform as \[\delta_{z,i} = \theta_i(x_i,t)\delta_i,\] 
and hence 
\[
V(x,\delta) = \sum_{i\in\N_{1,n}}\delta_i'\theta_i(x_i,t)\theta_i(x_i,t)\delta_i=|\delta_{z,i}|^2
\]
hence the CCM is Euclidean distance in $z$. Now, it was proven by \cite{manchester_control_2015} that the CCM conditions implying that energy is a CLF are invariant under smooth changes of coordinates, hence $|z-z^\star|^2$ is a CLF.
\end{pf}

\begin{rem}
	In the general case of vector nodes, one can still construct the differential change of coordinate
$\delta_{z,i}= \Theta_i(x_i,t)\delta_i$ by factoring $M_i(x_i,t)= \Theta(x_i,t)'\Theta(x_i,t)$ so that $|\delta_z|^2 = \delta'M(x,t)\delta$. However In general $\Theta$ is not integrable, so we may not be able to construct $z(x,t)$ with $\pder[z]{x} = \Theta(x,t)$.
\end{rem}

\section{Main Result}\label{sec:main}

An important sub-class of nonlinear networked systems is those with nonlinear local (node) dynamics, but positive (cooperative) linear coupling between nodes. The main result of this paper is to extend the result of \cite{khong_diagonal_2016} to this class of systems.

To be precise, consider a system with state $x\in\R^n$ and dynamics that decomposes as
\begin{equation}\label{eq:sys1}
\dot x_i = f_i(x_i,t)=g_i(x_i,t)+\sum_{j=1}^n k_{ij}(t)x_j, \quad i = 1, 2, ..., n,
\end{equation}
where $g_i,k_{ij}$ are piecewise-continuous in $t$, and $\pder[g_i]{x_i}$ is uniformly bounded above (in $x$ and $t$) for each $i$, and $k_{ij}\ge 0$ for $i\ne j$.

\begin{thm} \label{thm:main} Suppose a system of the form \eqref{eq:sys1} is contracting with respect to a metric of the form $V(x,\delta,t) = \delta'M(t)\delta$. Then it is contracting with respect to a sum-separable metric of the form 
\[
V_d(x,\delta,t) = \sum_i m_i(t)|\delta_i|^2.
\]
\end{thm}
\begin{pf}
For each $i\in \N_{1,n}$, define 
\[
\bar G_i(t) = \sup_{x_i} \pder[g_i(x_i,t)]{x_i}
\]
which, by assumption on $g_i$ is a piecewise-continuous function of $t$. Now, construct the diagonal matrix $\bar G(t)\in\R^{n\times n}$ with each $\bar G_i(t)$ as the $i^{th}$ 
diagonal elements, and consider the linear time-varying system
\begin{equation}\label{eq:comparison}
\dot z = (\bar G(t)+K(t))z
\end{equation}
where $K(t)$ is a matrix with $i,j$ element given by $k_{i,j}(t)$.

By assumption that $k_{ij}(t)\ge 0$ for $i\ne j$, this system is positive. We will now show that it is also exponentially 
stable. 

By definition, for each $i$ and $t$ there exists a sequence $\bar x_i^k(t)$ such that
$\pder[g_i]{x_i}(x_i^k(t),t)\rightarrow \bar G_i(t)$ as $k\rightarrow
\infty$. Now, construct the sequence of states $x^k(t) = [x_1^k(t), x_2^k(t),
  ..., x_n^k(t)]'$. By construction $\pder[g]{x}(x^k(t),t)\rightarrow \bar G(t)$ as
$k\rightarrow \infty$. 

Now, by assumption that the system is contracting, for every $x$ and $t$ and for some $\lambda>0$ and some uniformly bounded $Q(t)$, the following inequality holds
\[
\dot Q(t) + \pder[f]{x}(x,t)Q(t)+Q(t)\pder[f]{x}(x,t)\le - 2\lambda M(t).
\]
By continuity of $f$, we therefore have the same condition with $\pder[f]{x}(x,t)$ replaced by $\bar G(t)+K(t)$. This implies that the LTV system \eqref{eq:comparison} is exponentially stable and meets the conditions of Theorem \ref{thm:khong}, and therefore has a diagonal Lyapunov function.

Hence there exists a separable Lyapunov function $V(z,t)=z'M(t)z$ with $M(t)>0$ diagonal and \[\dot M+(\bar G+K)'M+M(\bar G+K)<-\nu I\] for all $t$, and hence by uniform boundedness \[(\bar G+K)'M+M(\bar G+K) \le -2\l_d M\] for some $\l_d>0$. 

Denote by $m_i$ the $i^{th}$ diagonal element of $M$.
By construction of $\bar G$ and the fact that $M_i>0$ we have
\[
M\pder[f]{x}-M(\bar G+K)=\sum_{i\in \N_{1,n}} M_i(\pder[g_i]{x_i}-\bar G_i)\le 0,
\]
from which it follows that
\[
\dot M(t)+\pder[f]{x}'(x,t)M(t)+M(t)\pder[f]{x}(x,t)\le 2\lambda_d M(t).
\]
This completes the proof of the theorem.
\end{pf}

%\section{Extension to Block Case}
%
%Same as above, but now with each $x_i\in\R^{n_i}$.
%
%Take $\bar G_i$ to be positive and satisfy $\|G_i\|=\sup_{x_i} \|\pder[g_i]{x_i}\|$. With respect to $M$?
%
%Contraction implies lossless S-procedure condition for diagonal metric.
%
%Converse: show existence of a sequence in $x$ and $\delta_x$
%	
%\section{Time-varying case}
%
%Utilizing the result of \cite{khong_diagonal_2016}, we can extend the above result to time-varying contracting system. Note that this includes the important case of systems with periodic inputs, and implies entrainment to a periodic solution \cite{Lohmiller98}.
%
%\begin{thm}
%Consider a system
%
%such that there exists a metric $M(t)$ satisfying $\alpha_1 I \le M(t)
%\le \alpha_2 I$ for some constants $\alpha_2\ge \alpha_1>0$, and
%\begin{equation}\label{eq:contractionTV}
%\dot M+\pder[f]{x}'M+M\pder[f]{x}+2\lambda M\le 0
%\end{equation}
%Then there exists a sum-separable contraction metric.
%\end{thm}
%\begin{pf}
%Let
% We have
%\[
%\dot M+(\bar G+K)'M+M(\bar G+K)\le -2\lambda M
%\]
%since contraction would fail otherwise for $\bar x,t$. It follows that
%the LTV system defined by
%\[
%\dot z = (\bar G(t)+K(t))z
%\]
%is positive and exponentially stable. Therefore, by Theorem \ref{thm:khong}, there exists a time-varying diagonal Lyapunov function.
%
%Then this is a contraction metric
%
%
%\end{pf}
%
%
\section{Additional Results}

In this section, we present some other conditions under which it is straightforward to show existence of a separable Lyapunov function.

\subsection{Local Existence Along Trajectories}

It is known that if a monotone system is exponentially stable at the origin (i.e. $\pder[f]{x}(0)$ is Hurwitz), then there exists sum and max separable Lyapunov functions in a neighbourhood of the origin \citep{dirr2015separable}. These results extend easily to contraction metrics, and in fact the following stronger condition is available.

\begin{thm}
	Assume system \eqref{eq:sys} is contracting and monotone. Then in a neighbourhood of any bounded solution $x_0(t), t\in[0,\infty)$ of \eqref{eq:sys} there exists a sum-separable contraction metric.
\end{thm}
\begin{pf}
For any bounded solution $x_0(t)$ we can construct a compact set $\mathcal X_0$ such that $x_0(t)\in\mathcal X_0$ for all $t$. furthermore, by continuity inside this set the elements of $\pder[f]{x}$ are uniformly bounded.

	Since the system is contracting and monotone, the LTV differential dynamics are exponentially stable and positive along any solution. 	
	Therefore, by Theorem \ref{thm:khong} there exists a diagonal Lyapunov function $\delta'M(t)\delta$ for the differential dynamics, which therefore satisfies the condition
	\[
	\dot M(t)+\pder[f]{x}(x_0(t),t)M(t)+M(t)\pder[f]{x}(x_0(t),t) \le -2\lambda M(t)
	\]
	for all $t\ge 0$.
	
	Now, since $M$ is uniformly bounded, by choosing some $\lambda_0<\lambda$ we have that for all $t\ge 0$,
	
	\[
	\dot M(t)+\pder[f]{x}(x,t)'M(t)+M(t)\pder[f]{x}(x,t) \le -2\lambda_0 M(t)
	\]
	for all $x$ in some neighbourhood of $x_0(t)$. This completes the proof of the theorem.
\end{pf}
\begin{rem}
	This does not imply that the neighbourhood on which the contraction condition holds is forward invariant, however as long as the diameter of the neighbourhood does not shrink to zero as time goes to infinity, such a set can be easily constructed.
\end{rem}
\begin{rem}
	Unlike the results of \citep{dirr2015separable}, this theorem applies even to systems with no equilibria, e.g. systems with periodic forcing.
\end{rem}

\begin{rem}
Note that by applying this result along {all} solutions we can construct lead to a {\em diagonal} metric of the form
\[
V(x,\delta) = \sum_i m_i(x)\delta_i^2.
\]
However, it is {\em  not} guaranteed that $m_i(x)$ depends only on $x_i$, which is the condition for true sum-separability of the metric.
\end{rem}

\subsection{Networks with Weakly Nonlinear Coupling: Small Gain Condition}                                                                         

He we briefly remark that a well-known construction of separable Lyapunov functions based on small-gain type coupling can be extended to the case of contraction and time-varying systems. 
The conditions we discuss date back to early results in decentralized control and vector Lyapunov functions \citep{moylan1978stability, sandell_survey_1978}, and have been substantially generalized recently, e.g.\cite{rueffer_small-gain_2010, ito_capability_2012}. A similar construction was considered in \cite{russo2013contraction}.

For the results of this section, the node states can have arbitrary dimensions. Additionally, there is no assumption that the original system is monotone, though a linear time-varying comparison system is.

Suppose  for some system of the form \eqref{eq:sys} with $N$ nodes of dimension $n_i$, one constructs local differential storage functions $V(x_i,\delta_i,t)$, e.g. of the form $V(x_i,\delta_i,t) = \delta_i'M(x_i,t)\delta_i$.
Suppose it can be verified that
\[
\dot V_i \le \sum_{j\in N_i} \alpha_{ij} V_j
\]
where $\alpha_{ii}<0$ and $\alpha_{ij}\ge 0, i\ne j$.
Note that unlike \cite{russo2013contraction} this allows nonlinear and time-varying gains between subsystems, due to the possible state-dependence of $M(x,t)$.

%This implies the class $\kappa$ function representation?

Now, consider the $N\times N$ matrix
\[
H = \begin{bmatrix}
	\alpha_{11} & \alpha_{12} & \hdots & \alpha_{1n}\\
	\alpha_{21} & \alpha_{22}  & \hdots & \alpha_{2n}\\
	\vdots & & \ddots & \vdots \\
	\alpha_{n1} & \alpha_{n2}  & \hdots & \alpha_{nn}
	\end{bmatrix}
\]
and the the system
\[
\dot z = Hz
\]
is a positive LTI system. Hence, if this system is exponentially stable, there exists  a linear Lyapunov function on the positive orthant
\[
V_1(z)= \sum_i p_i z_i.
\]
which verifies exponential decrease.
So, constructing the metric
\[
V(x,\delta) = \sum_i p_i V_i(x_i,\delta_i)
\]
it is clear that this is a sum-separable contraction metric for the original system.

\begin{rem}
There also exists max-type Lyapunov function  $V_2 = \max_i\{q_i z_i\}$ for the linear comparison system, which could be used to construct a max-separable contraction metric.
\end{rem}

\subsection{Networks with Weakly Nonlinear Coupling using the S-Procedure}

A disadvantage of the results of the previous section is that they depend heavily on knowing a good construction of the local storage functions $V_i$. In this section, we give results that are based only on explicit bounds on the vector field.

The key result can be considered a differential version of the results in \cite{colombino2015robust}, based on the losslessness of the S-Procedure for positive systems. Related results were established in \cite{tanaka2013symmetric}.

In particular, consider systems with scalar nodes of the form 
\[
\dot x_i = f_i(x_i)=g_i(x_i)+\sum_{j=1}^{j=n} k_{ij}x_j+h_i(x,t),  \quad i\in \N_{1,n}
\]
where $k_{ij}\ge 0$ and $\pder[g_i]{x_i}$ is uniformly bounded, and the functions $h_i$ satisfy  
\begin{equation}\label{eq:hbound}
\delta_x'\pder[h]{x}'\pder[h]{x}\delta_x\le \psi^2\delta_x'H'H\delta_x
\end{equation}
for all $x, \delta_x$,
for some non-negative matrix $H$ and some $\psi\in\R$.

\begin{thm}
	Suppose the system is contracting for all nonlinear functions $h_i$ satisfying \eqref{eq:hbound} as verified by a metric $V(x,\delta)=\delta'M\delta$, then $M$ can be taken to be diagonal.
\end{thm}
\begin{pf} We skip some details since they are similar to Theorem \ref{thm:main} and \cite[Theorem 4]{colombino2015robust}.

Construct the matrix $\bar G$ as in the proof of Theorem \ref{thm:main} and consider the system
\[
\dot z = \left(\bar G + K+\pder[h]{x}(x,t)\right)z
\]
where $h$ is a vector function with elements $h_i$. As in Theorem \ref{thm:main}, a metric proving contraction for this system also proves contraction for the true nonlinear system.

Now, by \cite[Theorem 4]{colombino2015robust}, stability of this system for all $h_i$ satisfying \eqref{eq:hbound} is equivalent to the existence of a {\em diagonal} matrix $M$ for which
\[
\begin{bmatrix}
	\bar G'M+ M\bar G& M\\M & 0
\end{bmatrix}
+\sum_i \theta_i
\begin{bmatrix}
	 \psi^2H_i'H_i\\0 & -\Pi_i
\end{bmatrix}<0.
\]	
It is straightforward to show that this same $M$ leads to a sum-separable contraction metric.
\end{pf}

\subsection{Stability verification via a virtual system}

Given a system \eqref{eq:sys}, a {\em virtual system} is another system driven by $x$:
\[
\dot y = \bar f(y,x,t)
\]
with the special property that $\bar f(x,x,t) = f(x,t)$ for all $x,t$. Analysing such systems is common in observer design and synchronisation problems, but can also be used to establish stability of other classes of systems \citep{wang2005partial,jouffroy2004methodological}.

Let us detail a particular example of this application which can make use of positive systems theory. Many networked dynamical systems can be factored as
\begin{equation}\label{Nxx}
\dot x = N(x,t)x
\end{equation}
where $N(x,t)$ is a positive matrix for all $x,t$. This system has the origin ($x(t)=0 \forall t$) as a solution. To establish stability of this solution we can introduce the virtual system
\begin{equation}\label{eq:virtual}
	\dot y = N(x(t),t)y
\end{equation}
where $x(t)$ is the solution of the true system but considered as time-variation for the virtual system. Suppose this system is exponentially stable for all $x(t)$, and note that a particular solution of \eqref{eq:virtual} is $y(t)=x(t)$ for all $t$. and hence solutions of \eqref{Nxx} converge to zero exponentially.

Now, since \eqref{eq:virtual} represents a family of positive LTV systems parameterized by the initial condition $x(0)$ for each solution $x(t)$,  it follows from \cite{khong_diagonal_2016} that there exists a family of diagonal Lyapunov function
\[
V(y,t) = \sum m_i(t,x(0)) |y_i|^2
\]
 verifying stability of all solutions $y(t)$ of \eqref{eq:virtual}. Therefore, establishing stability by way of a positive virtual system is simplified compared to the general case.

\section{Conclusions}

In this paper we have investigated the problem of when a contracting system has a {\em separable} contraction metric. This problem is motivated by the need to find simplified stability conditions for large-scale nonlinear systems, and also be recent work that shows that nonlinear distributed control design problems can be made convex if a sum-separable contraction metric exists.

  To our knowledge a complete characterisation remains unknown, however in this paper we have demonstrated a number of simple cases cases in which it can be established.

%Forni and Sepulchre, differentially positive systems?

\bibliographystyle{ifacconf}
\bibliography{monotone.bib}

\end{document}